\documentclass[journal,article,submit,pdftex,moreauthors]{mdpi}

\firstpage{1} 
\pubvolume{1}
\issuenum{1}
\articlenumber{0}
\pubyear{2022}
\copyrightyear{2022}
\datereceived{} 
\dateaccepted{} 
\datepublished{} 

\hreflink{https://doi.org/} % If needed use \linebreak
\pdfoutput=1

\usepackage{subcaption}
\usepackage{csquotes}
\usepackage{graphicx}

\Title{Compared analysis of DInSAR data from ascending and descending orbits of Sentinel-1: the Cazzaso case study}

\TitleCitation{Title}

\Author{Giuseppe Buono $^{1, 2}$, Paolo Facchi $^{1, 2}$, Raffaele Nutricato $^{3}$,  Luciano Guerriero $^{4}$, Francesco Vincenzo Pepe $^{1, 2}$,  Cosmo Lupo $^{4,2}$ and Saverio Pascazio $^{1,2}$}

\AuthorNames{Giuseppe Buono, Raffaele Nutricato, Paolo Facchi, Luciano Guerriero, Francesco Vincenzo Pepe, Cosmo Lupo and Saverio Pascazio}

\AuthorCitation{Lastname, F.; Lastname, F.; Lastname, F.}

\address{%
$^{1}$ \quad Dipartimento Interateneo di Fisica, Universit\`a  di Bari, I-70126 Bari, Italy\\
$^{2}$ \quad  Istituto Nazionale di Fisica Nucleare, Sezione di Bari, I-70126 Bari, Italy\\
$^{3}$ \quad Geophysical Applications Processing s.r.l., I-70126 Bari, Italy\\
$^{4}$ \quad Dipartimento Interateneo di Fisica, Politecnico di Bari, I-70126 Bari, Italy\\}

\corres{Correspondence: giuseppe.buono1@uniba.it}

\abstract{Differential SAR interferometry (DInSAR), by providing displacement time series over coherent objects on the Earth's surface (persistent scatterers), allows to analyze wide areas, identify ground displacements, and study their evolution at large times. In this work we implement an innovative approach that relies exclusively on line-of-sight displacement time series, applicable to cases of correlated persistent-scatterer displacements. We identify the locus of the final positions of the persistent scatterers and automatically calculate the lower bound of the magnitude of the  potential three-dimensional displacements. We present the results obtained by using Sentinel-1  data for investigating the ground stability of the hilly village Cazzaso located in the Italian Alps (Friuli Venezia Giulia region) in an area affected by an active landslide. SAR datasets acquired by Sentinel-1 from both ascending and descending orbits were processed using the SPINUA algorithm. Displacement time series were analysed in order to solve phase unwrapping issues and displacement field calculation. }


\begin{document}

%%%%%%%%%%%%%%%%%%%%%%%%%%%%%%%%%%%%%%%%%%
\setcounter{section}{0} %% Remove this when starting to work on the template.

\section{Introduction}
 
Differential SAR interferometry (DInSAR) techniques provide displacement time series over coherent objects on the Earth's surface (Persistent Scatterers - PS), enabling the analysis of wide areas to identify ground displacements, and the study of evolution phenomena at decadal time scales~\cite{Ferretti_2000, Ferretti_2001, Colesanti_2003}. These techniques are among the most promising and reliable ones in the detection and monitoring of slope instabilities. Landslides are among the most difficult types of ground displacements that can be investigated through remote sensing techniques, as their velocity rates may cover several orders of magnitude, from very low creeping to sudden catastrophic failures, thus pushing detection performance to its limits. 

Further challenges related to the application of Multi-Temporal SAR Interferometry (MTInSAR) analysis for slope instability monitoring concern: (i) satellites-PS system geometry for identifying complex kinematics, and (ii) advanced time series analysis required for data correction and phase unwrapping procedures. 

In this work we present an innovative approach that relies exclusively on line-of-sight (LOS) displacement time series, applicable to cases of correlated persistent-scatterer displacements. 
By combining displacement time series from both ascending and descending orbits of Sentinel-1 (S1) we investigated the ground stability of Cazzaso village located in the Italian Alps (Friuli Venezia Giulia region)~\cite{Zuliani_2022,Tunini_2024}. SAR datasets acquired by S1 from both ascending and descending orbits were processed by using the SPINUA MTInSAR algorithm, in order to exploit the potential of this satellite mission to investigate ground displacements related to the slope instabilities of a test site characterized by a high PS density~\cite{Bovenga_2004}. The high density of PS distributed thorough Cazzaso village test site and the high correlation among their time series allow us to identify the locus of the PS final positions and automatically calculate a lower bound to the magnitude of the three-dimensional displacements, providing useful insight about displacement field calculation. 
  
%%%%%%%%%%%%%%%%%%%%%%%%%%%%%%%%%%%%%%%%%%
\section{Materials and Methods}

\subsection{Test Sites Description and Datasets}
\begin{figure}[H]
\centering
\includegraphics[width=0.7\textwidth]{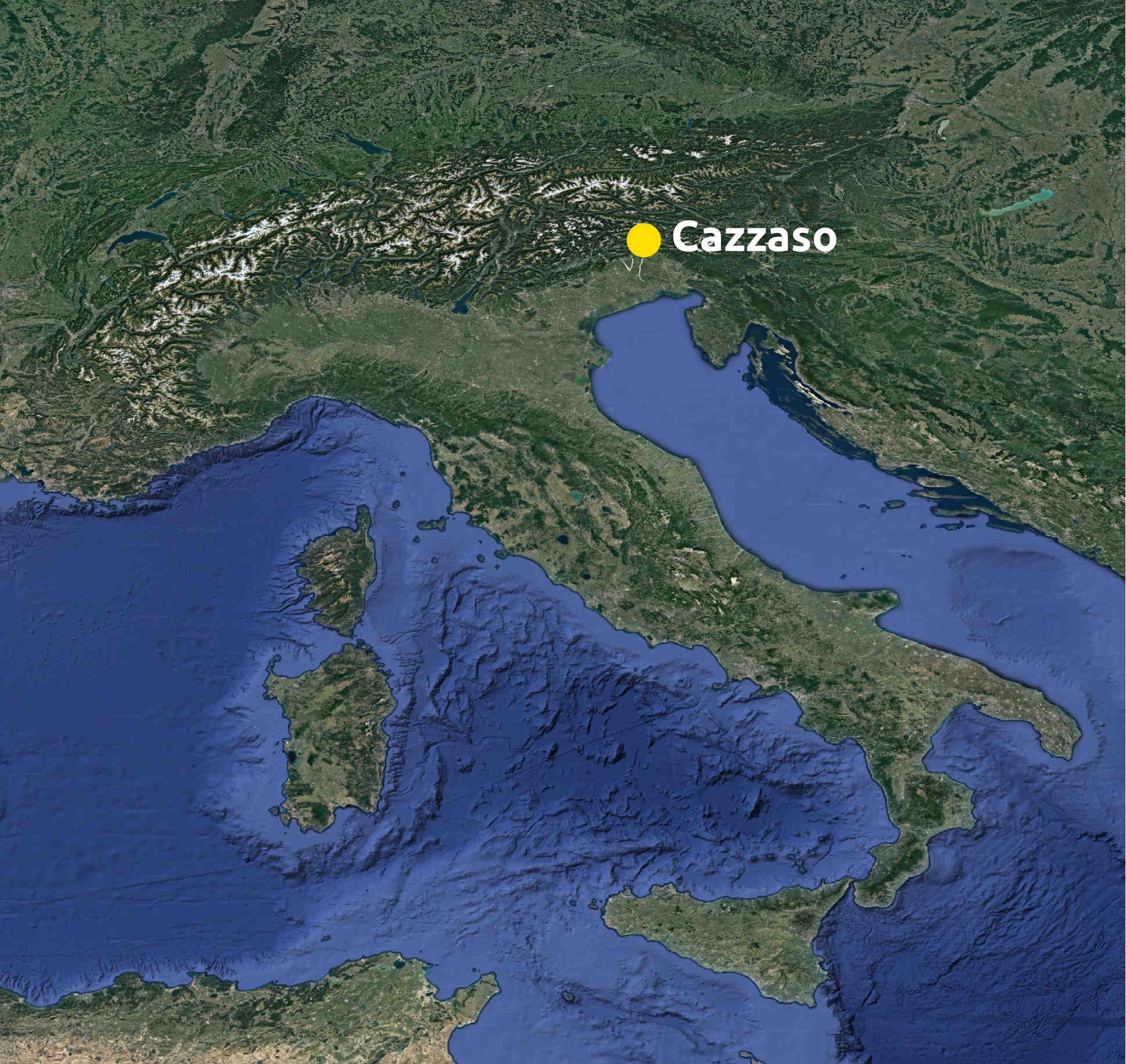}
\caption{Location of Cazzaso village in the Italian Alps.}
\label{fig1}
\end{figure}

Cazzaso is a village located in Northern Italy, in the eastern portion of the Italian Alps (Figure~\ref{fig1}) (Coordinates: Latitude 46.432$^o$ N, Longitude 12.995$^o$ E). Cazzaso is situated in a region that has been subjected to an active landslide for years and is therefore under constant monitoring. We chose Cazzaso as a specific case study because it is an exceptionally complex site, to the extent that the European Ground Motion Service (EGMS) does not report any measurement points in this area. This is despite the fact that data processing was carried out using multiple algorithms, harmonized by four leading global players in the field and validated by the European Environment Agency (EEA). Even with this rigorous approach, the analysis remained particularly challenging, highlighting the site's complexity and the limitations of conventional monitoring techniques.

\begin{table}[H] 
\caption{Dataset characteristics.\label{tab1}}
\newcolumntype{C}{>{\centering\arraybackslash}X}
\begin{tabularx}{\textwidth}{cCCC}
\toprule
\textbf{Test Site}	& \textbf{Sensor} & \textbf{Geometry} & \textbf{Time Interval} \\
\midrule
Cazzaso		& Sentinel-1  & Ascending  & 01/01/2017-21/06/2021\\
 			&	  & Descending & 01/01/2017-21/06/2021\\
\bottomrule
\end{tabularx}
\end{table}

Table \ref{tab1} schematically reports the datasets used in the present work. S1 data are acquired in interferometric wide-swath (IW) mode, with a ground pixel size of about $5 \times 20$ m$^{2}$ (range $\times$ azimuth).
Data series cover a time period from 2017 to 2021, those available at the time when the DInSAR processing was carried out. The results derived from the available datasets are suitable for the time series analysis we present in the next section, and perfectly suit the aim of this work.
All data series were processed through the SPINUA MTIn-SAR processing suite~\cite{Bovenga_2004}, analyzing a database of PS. Any record includes the coordinates of the PS, the average value of the temporal coherence, the average velocity projected along the line-of-sight (VLOS), and two complete time series with values calculated at each acquisition date: one for the relative LOS displacements and one for the radar cross section amplitude.
The temporal baseline depends on the number of satellites in orbit. In this case, during the first few months of the time series, data were collected every 12 days. Then, after the launch of a second satellite, the temporal baseline was reduced to 6 days, where it remained constant until the end of the analyzed period.

%%%%%%%%%%%%%%%%%%%%%%%%%%%%%%%%%%%%%%%%%%
\section{Results}

\subsection{Cazzaso Test Site}
\begin{figure}[H]
\centering
\includegraphics[width=0.7\textwidth]{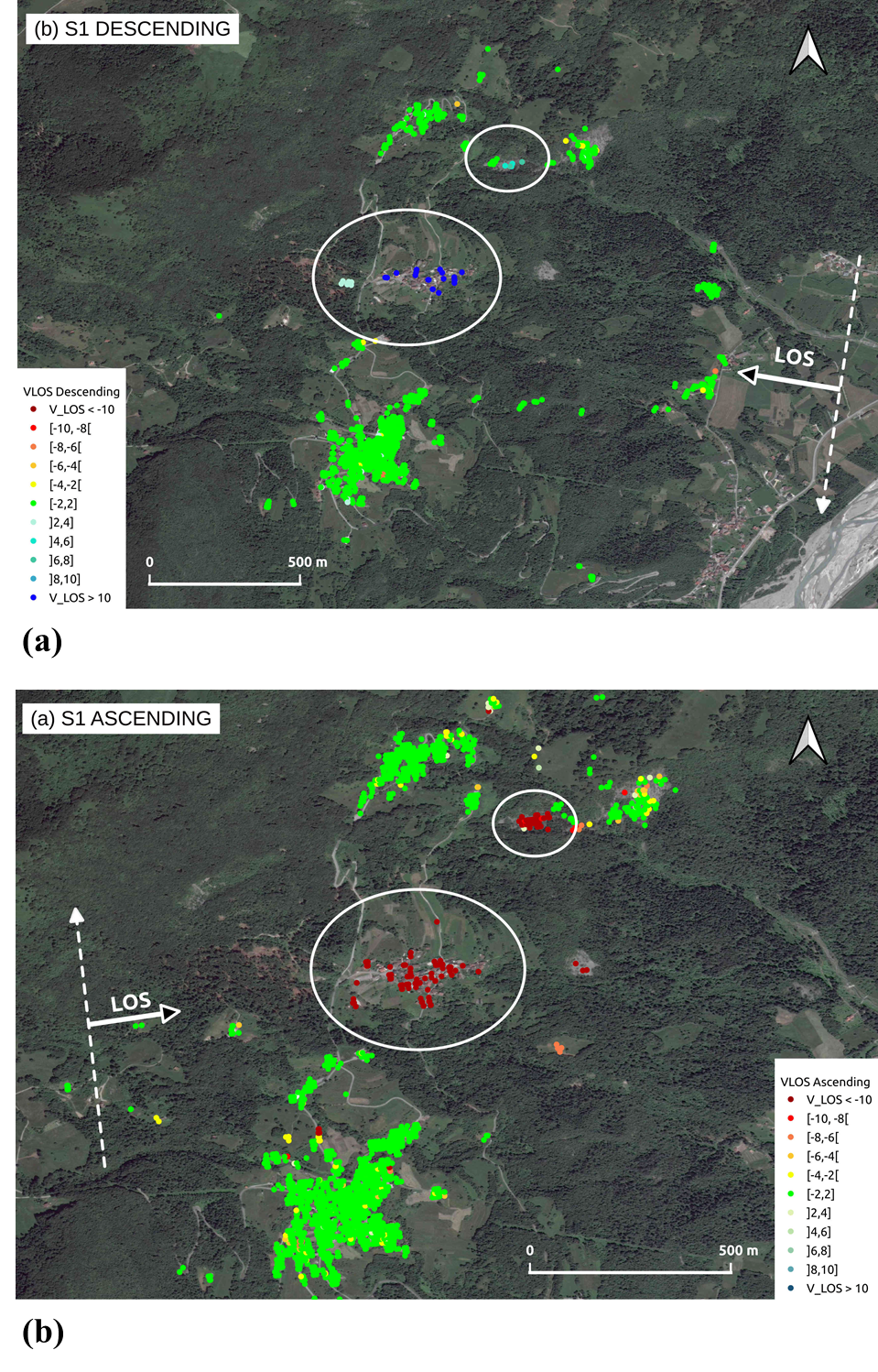}
\caption{Observation area around the village of Cazzaso with the detected persistent scatterers. Both figures show the satellite movement direction (dashed lines) and the line-of-sight (LOS). The colour of the persistent scatterers depends on the value of the VLOS (average velocity projected along the LOS). Green points represent relatively stable PS with low displacement rates. Points in shades of red indicate displacement rates moving away from the satellite. Points in shades of blue represent displacement rates moving towards the satellite.}
\label{f_002}
\end{figure}

\begin{figure}[H]
\centering
\includegraphics[width=0.7\textwidth]{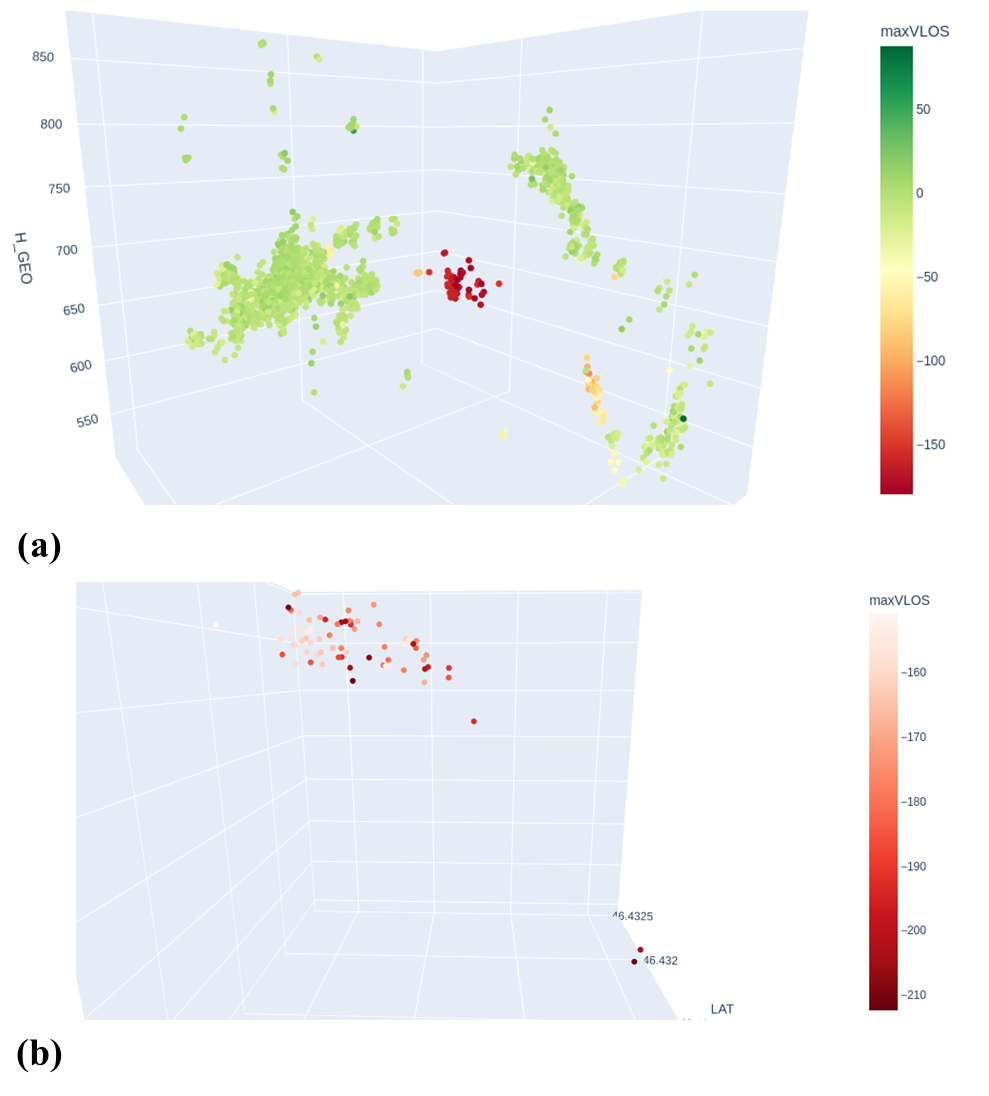}
\caption{3D rendering of the persistent scatterers (ascending direction). Green points represent relatively stable PS with low displacement rates. Points in shades of red indicate displacement rates moving away from the satellite. a) PS positions across the entire test site; b) A zoom in of the red points in a), from a  slightly different perspective, highlights the PS with the highest displacement rates, many of which are located in the village area. The image also shows the village situated on the edge of a hill, with additional PS visible downhill from the village.
}
\label{f_003}
\end{figure}

\begin{figure}[H]
\centering
\includegraphics[width=0.7\textwidth]{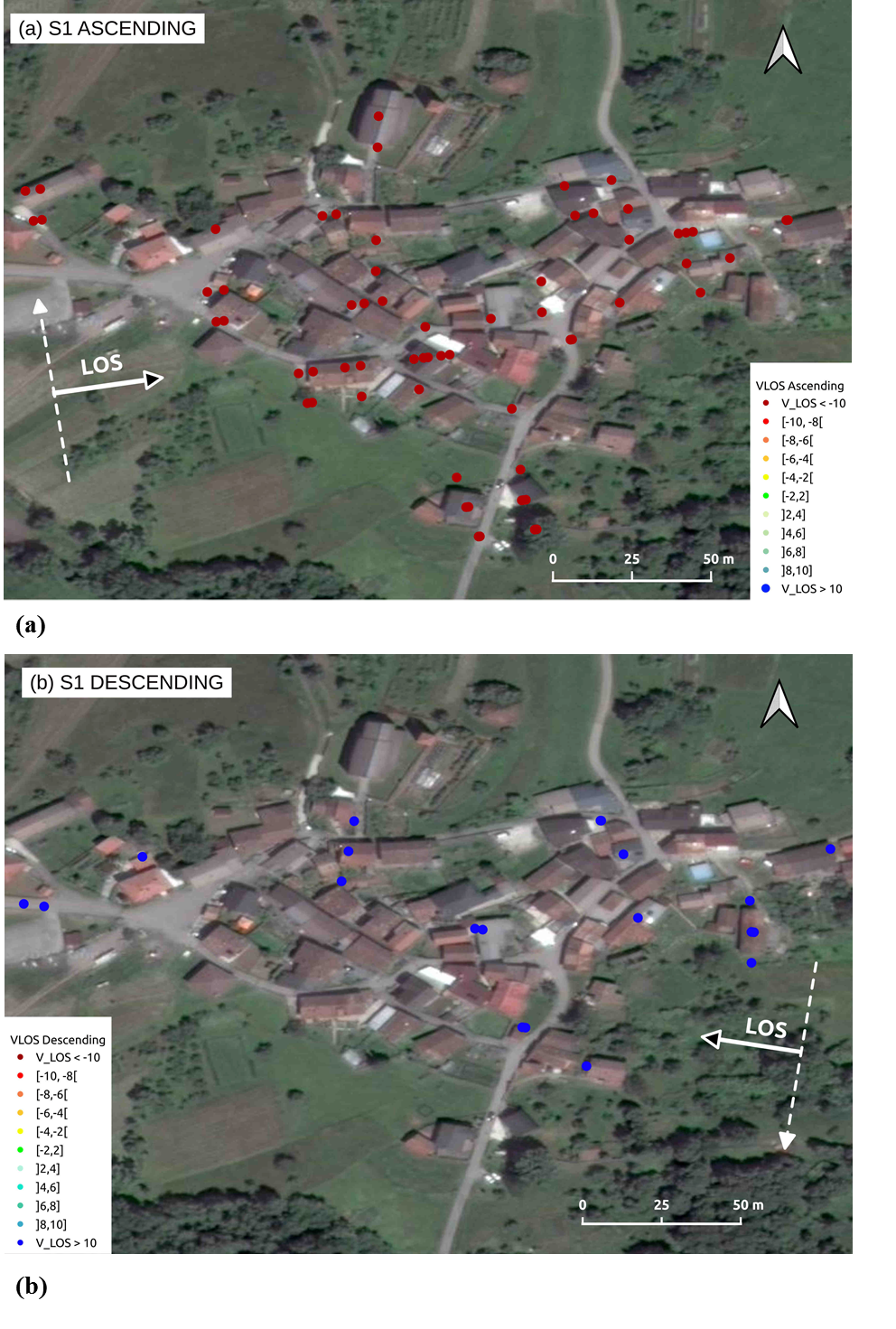}
\caption{Cazzaso Village PS:  a) Ascending;  b) Descending. Dashed arrows represent orbit directions. The ground projection of the LOS vector is also shown. Points in shades of red indicate displacement rates moving away from the satellite. Points in shades of blue represent displacement rates moving towards the satellite.}
\label{f_004}
\end{figure}

Figure \ref{f_002} shows the observation area around the village of Cazzaso, along with the persistent scatterers identified using the SPINUA algorithm to process the S1 ascending and descending datasets for the test site. Specifically, the PS detected by the satellite in its ascending and descending phase are shown  in Fig.~\ref{f_002}a and~Fig.~\ref{f_002}b, respectively. For convenience, PS registered by ascending (descending) orbit measurements will be called "ascending" ("descending") PS. Henceforth, time series, unless explicitly specified, will refer to relative LOS displacements measurements.
In both images, the satellite movement directions and the LOS, which is perpendicular to the satellite's heading direction, are indicated. The color of the PS points, represented on a color scale, indicates the VLOS, the average displacement velocity of the persistent scatterers projected along the satellite's line of sight, over the entire observation period:
\begin{itemize}
\item[$-$]Green points represent relatively stable PS with low displacement rates (< 2 mm/year);
\item[$-$]Points in shades of red indicate displacement rates moving away from the satellite;
\item[$-$]Points in shades of blue represent displacement rates moving towards the satellite.
\end{itemize}
The PS can be grouped into clusters that share some common characteristics, such as similar displacement rates or spatial proximity, highlighting patterns in the data. In this context, it was quite easy to cluster persistent scatterers based only on their geographical position. By looking at the figures, it is clear that  it is possible to identify clusters in the area even by eye.
In particular, it can be observed that there are two areas where PS with high displacement rates are more numerous: in the ascending direction, these are mainly red points, while in the descending direction, they are blue points. We will make use of this ``collective'' information in order to (conservatively) validate the displacements.
\pagebreak[4]
From a 3D rendering of PS positions, it is possible to observe that the two highlighted areas are disposed on two different sides of the same hill (Figure~\ref{f_003}a red and orange markers). The village lies on a steep surface, at the border of the "knee" of the hill profile, just before a sharp increase of the slope.
The central main area in Fig.~\ref{f_002}, marked with a circle, corresponds to the urban area of Cazzaso village (in Fig.~\ref{f_004} the village area is expanded). It is possible to observe 72 ascending PS and 20 descending PS in about $\text{0.054 km}^2$. Since the measurement resolution is influenced by the angle between the LOS and the target surface, in this test site, in ascending orbit measurements a higher number of PS is available than in descending ones.
The high density of PS, both in ascending and descending directions, equally distributed through the whole village, allows us to study the system as a whole, focusing on its interconnections. The following analysis will focus on this specific area.   
 
\subsection{PS time series} \label{ps_time_series}
\begin{figure}[H]
\centering
\includegraphics[width=0.8\textwidth]{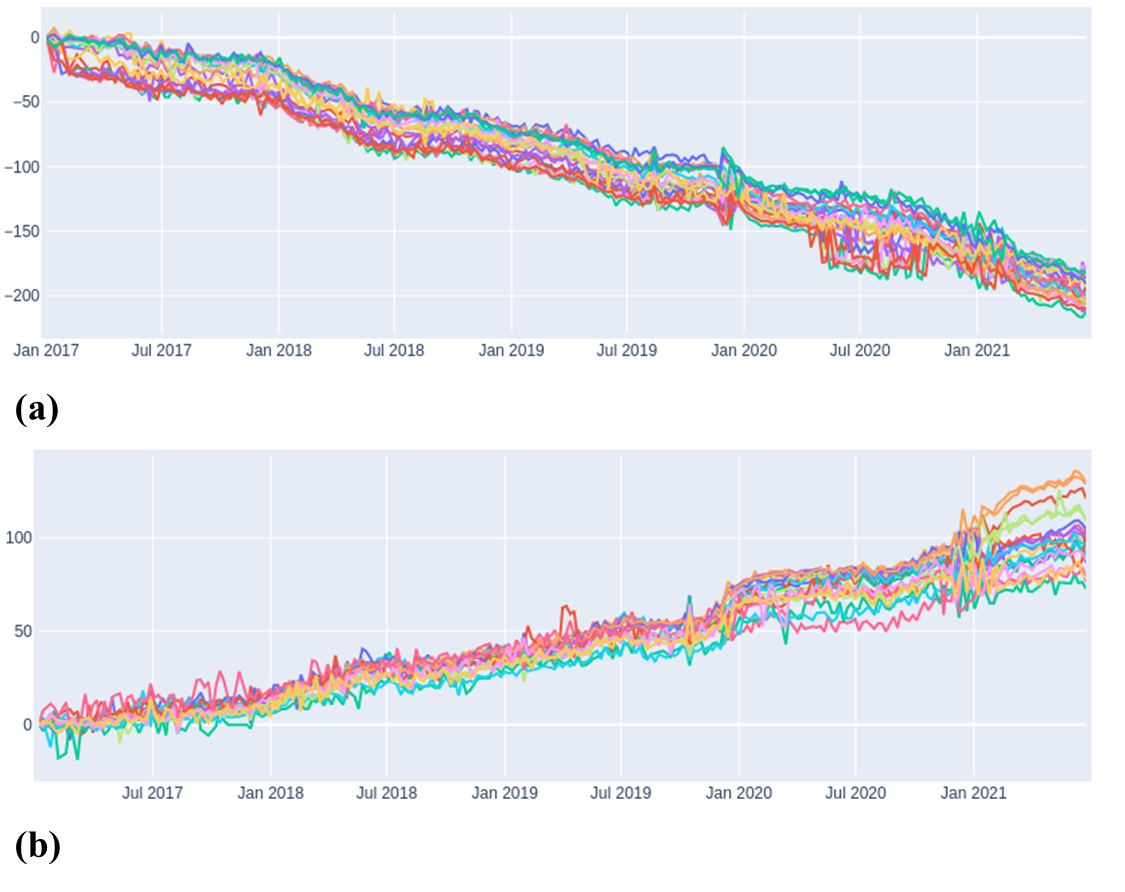}
\caption{PS displacements time series, for about 70 PS. The relative motion between the satellite and the persistent scatterer determines the sign of the displacements measured. a) Displacements of the persistent scatterers, as measured by the satellite during its ascending phase. b) Displacements of the persistent scatterers, as measured by the satellite during its descending phase.}
\label{f_005}
\end{figure}

\begin{figure}[H]
\centering
\includegraphics[width=0.8\textwidth]{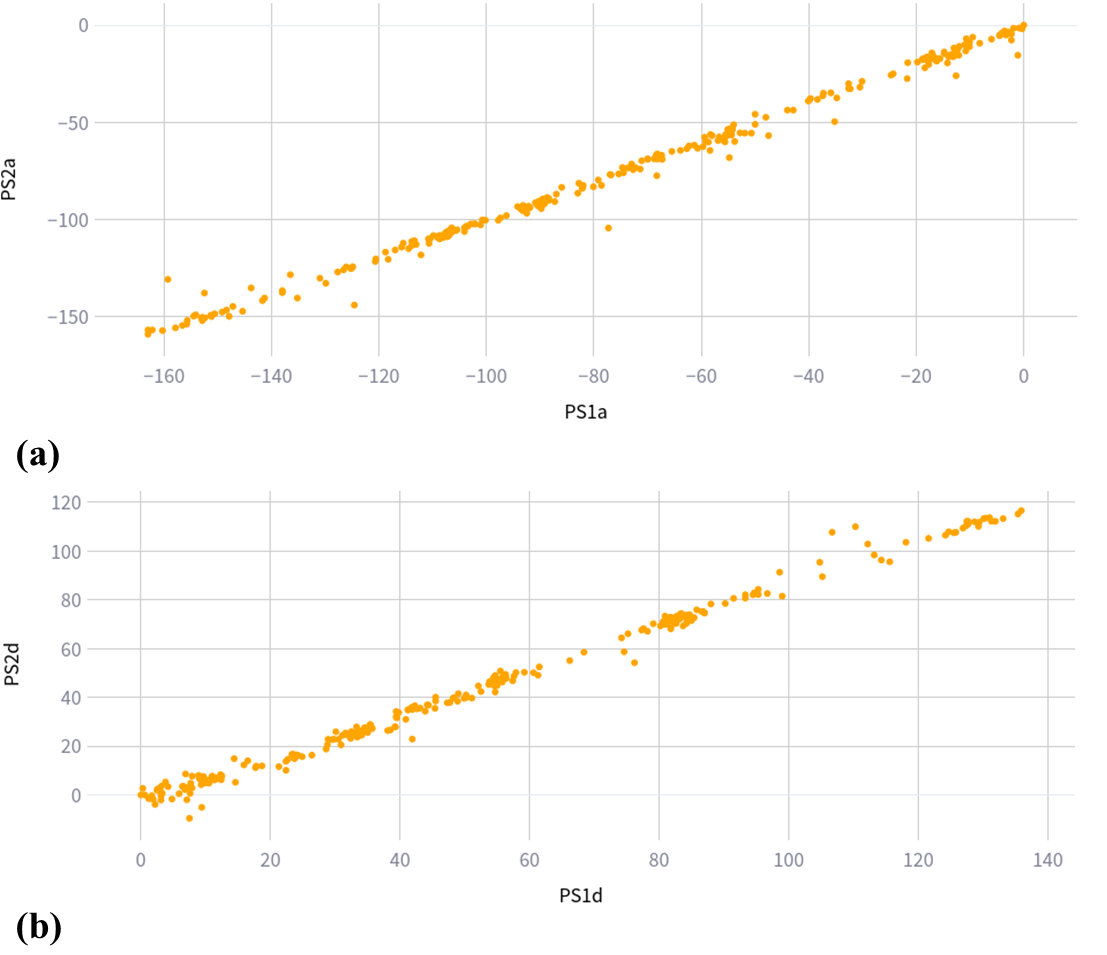}
\caption{Scatter plot of two displacement time series, both in ascending or descending mode, corresponding to two different PS in each image. The axes represent the LOS displacement of the two PS. The linear trend highlights the strong correlation between their displacements: a) ascending time series: the negative values of displacements mean that the PS are moving away from the satellite; b) descending time series: the negative values of displacements mean that the PS are moving towards the satellite.}
\label{f_006}
\end{figure}

By analyzing the time series of the PS in the village area and plotting them on the same graph, a strong correlation between the curves can be observed (see Fig.~\ref{f_005}). In addition, scatter plots comparing two different ascending PS time series (Fig.~\ref{f_006}a) or two different descending PS time series (Fig.~\ref{f_006}b) reveal a strong linear relationship in both cases. This correlation highlights an additional common characteristic shared by the PS within the same cluster, further supporting the conclusion that this area can be considered as a single cluster. PS time series show that in five years and a half, for all the ascending PS, the total LOS displacement is lower than -150mm, even \emph{without} considering possible phase unwrapping corrections which could further \emph{increase} the absolute value of the displacement (Figure~\ref{f_005}a). For all descending PS, the total LOS displacement is greater than 80mm, even without phase unwrapping enhancement (Figure~\ref{f_005}b). 
The average PS LOS displacement values during the whole observation period (over the PS in Fig.~\ref{f_004}), calculated before applying phase unwrapping enhancement procedures, are reported in Table~\ref{tab2}.

\begin{table}[H] 
\caption{PS total LOS displacement mean and standard deviation, derived from the cumulative displacements observed over the entire observation period, before phase unwrapping enhancement.\label{tab2}}
\newcolumntype{C}{>{\centering\arraybackslash}X}
\begin{tabularx}{\textwidth}{cCCC}
\toprule
\textbf{Direction}	& \textbf{Mean [mm]} & \textbf{Standard Dev [mm]} & \textbf{Time Interval} \\
\midrule
Ascending		    & -177  & 23  & 01/01/2017-21/06/2021\\
Descending			&  104	  & 17 & 01/01/2017-21/06/2021\\
\bottomrule
\end{tabularx}
\end{table}

\subsection{Phase unwrapping enhancement}
As previously mentioned, the phase unwrapping problem is particularly challenging for the Cazzaso site, despite being addressed using different procedures and algorithms. Even with the use of multiple procedures, harmonized by leading global players and validated by the EEA, the complexity of the site continues to pose significant difficulties, further emphasizing the limitations of conventional techniques in such demanding conditions. In the absence of  any other measurement reference (i.e. GPS measurements, other satellites data), S1 measurements are the only  data available to perform phase unwrapping. 

S1 works using a microwave sensor at wavelengths of 55.466mm. Considering the well known formula to convert phase measurements to distances,
\begin{equation}
d = \frac{\varphi \lambda}{4\pi},
\end{equation}
a $2\pi$ phase jump corresponds to a displacement gap of 27.7mm. Temporal data for a single PS only captures the phase evolution at that specific point. When dealing with persistent scatterers that exhibit extremely high noise, the process of phase unwrapping becomes particularly challenging~\cite{Chang_2016}. Quoting Ref.~\cite{Raspini_2018}, \enquote {It also worth recalling that TS (time series) are a zero-redundancy product, i.e., they contain one deformation measure per each SAR acquisition and for this reason, they are particularly sensitive to the phase noise.}. 

This long-standing problem will be solved on the basis of time series analysis supported by information gathered by:
\begin{itemize}
\item	collective considerations as PS relative positions;
\item	comparing time series of same direction (ascending or descending) PS;
\item	combining ascending and descending PS time series analysis.
\end{itemize}

Instead of relying solely on temporal data for a single PS, a network of PS can be used to provide additional context and resolve ambiguities through spatial correlation. The first step in spatial phase unwrapping is to cluster persistent scatterers into coherent and meaningful sets. Scatterers that are geographically close to one another are more likely to share similar environmental conditions. This proximity ensures that the differences between these points are smaller and more predictable. However, spatial proximity alone is not enough. A second layer of clustering involves examining the deformation time series of the scatterers. Displacement time series of persistent scatterers in the urban area of Cazzaso village show strong linear relationships and common trends that have been used in the phase unwrapping enhancement procedure sketched in Sec. \ref{ps_time_series}. A detailed account of such procedure will be discussed elsewhere.

\begin{figure}[hb!]
\centering
\includegraphics[width=0.8\textwidth]{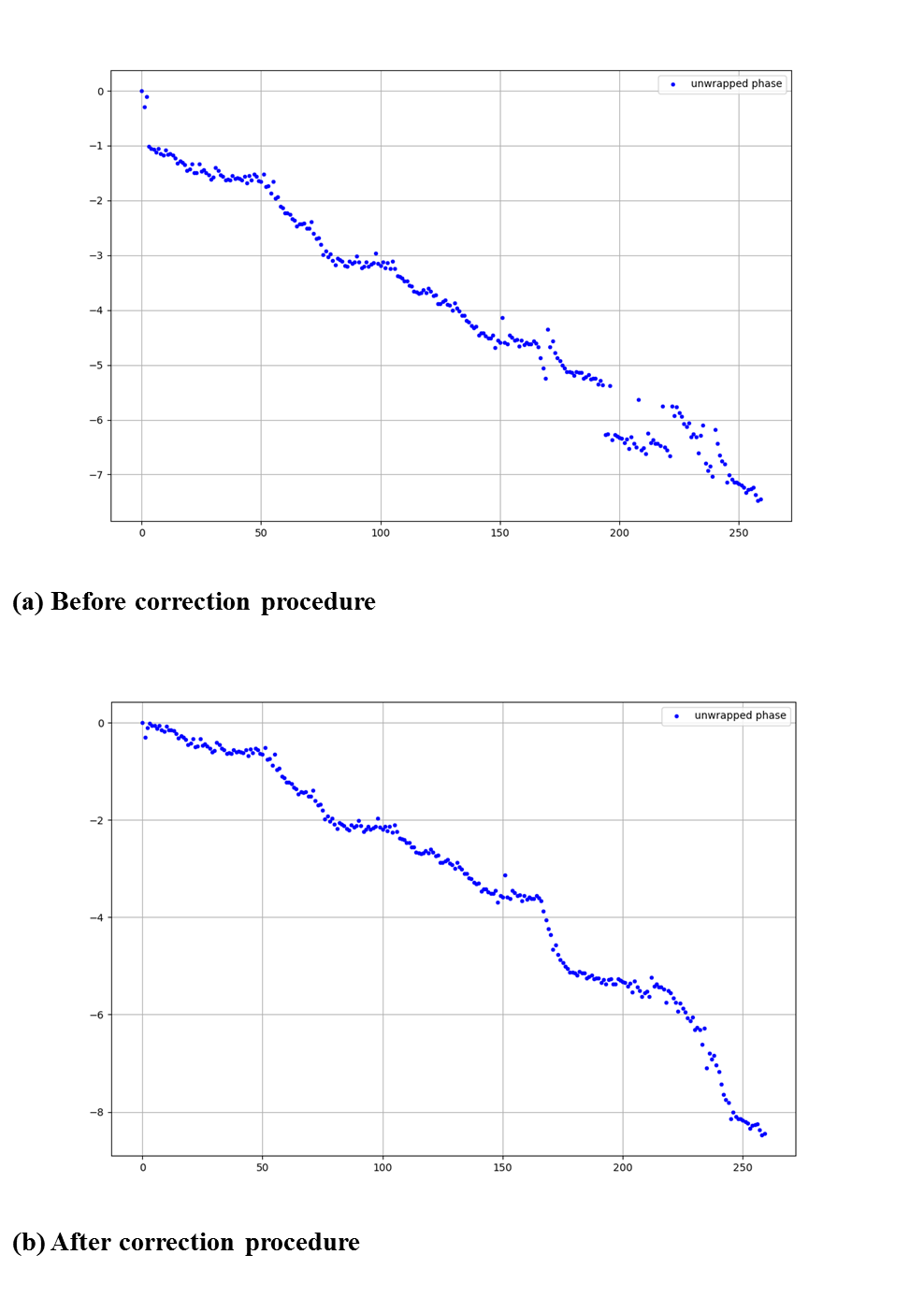}
\caption{Example of a persistent scatterer displacement time series. The diagram is normalized to the phase 2$\pi$ gap, corresponding to 27.7mm displacement for Sentinel-1 sensor. (a) before correction; (b) after correction.}
\label{f_007}
\end{figure}

\begin{figure}[hb!]
\centering
\includegraphics[width=0.8\textwidth]{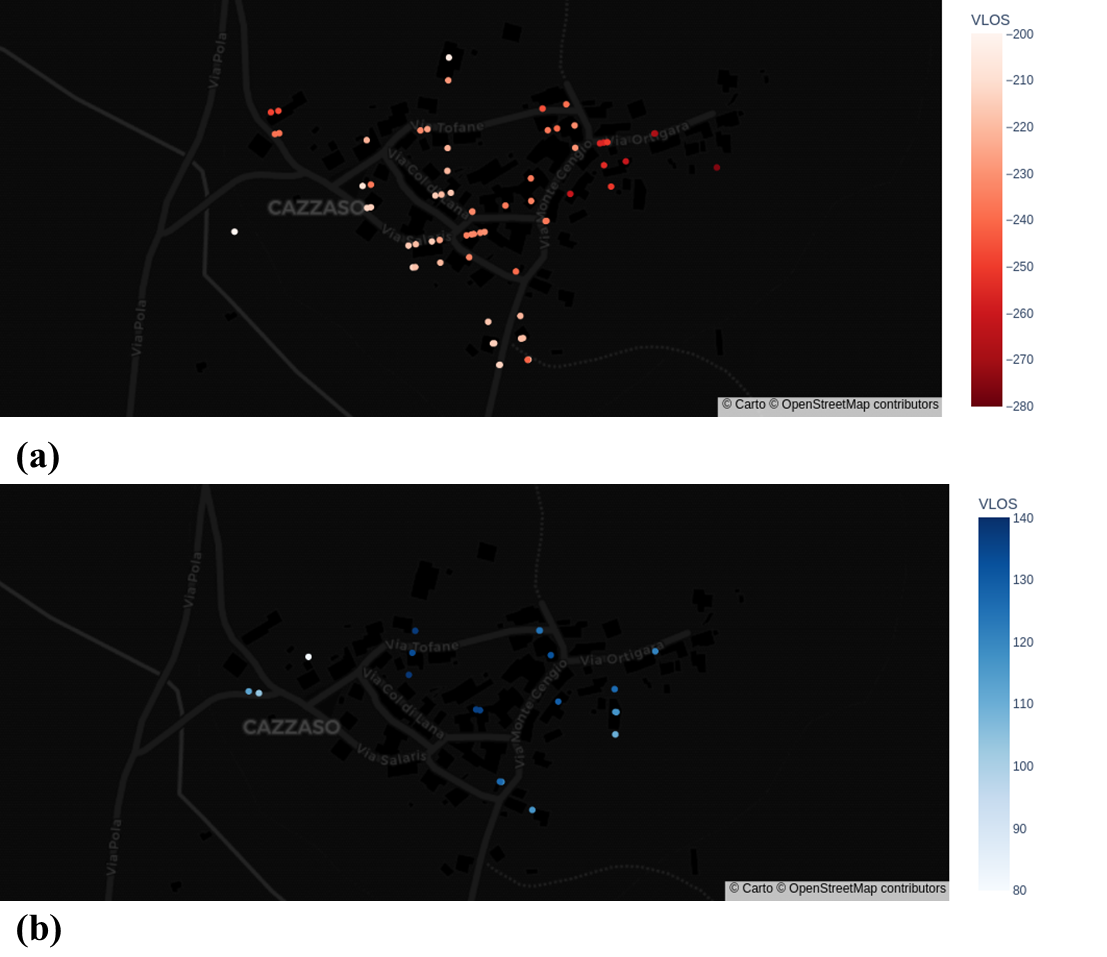}
\caption{PS LOS displacement gradient, calculated from the cumulative displacements observed over the entire observation period, after phase unwrapping enhancement: a) ascending phase; b) descending phase.
}
\label{f_008}
\end{figure}

In Fig.~\ref{f_007}, an example of phase unwrapping enhancement is presented. Several phase jumps in the data were corrected, restoring the continuity of the curve, even in sections where the deformation rate experienced significant changes. In this specific example, a potential error on the total displacement of about 12\% was corrected.
In Table \ref{tab4}, the average values of total LOS displacement after correction are reported. Comparing these values with those obtained before the correction (Table \ref{tab2}) reveals an average change greater than 20\% for ascending PS.
\begin{table}[H] 
\caption{PS total LOS displacement mean and standard deviation, derived from the cumulative displacements observed over the entire observation period, after phase unwrapping enhancement.\label{tab4}}
\newcolumntype{C}{>{\centering\arraybackslash}X}
\begin{tabularx}{\textwidth}{cCCC}
\toprule
\textbf{Direction}	& \textbf{Mean [mm]} & \textbf{Standard Dev [mm]} & \textbf{Time Interval} \\
\midrule
Ascending		    & -229  & 23  & 01/01/2017-21/06/2021\\
Descending		&  116	  & 11 & 01/01/2017-21/06/2021\\
\bottomrule
\end{tabularx}
\end{table}
At the end of the spatial unwrapping procedure, another interesting result was observed. While the PS before phase unwrapping correction appeared geographically disorganized, the enhancement process revealed clear gradients in the displacement data, both in the ascending and descending phases (Fig.~\ref{f_008}). These gradients only became visible after improving the unwrapping, and this information was then used to better understand the geological phenomenon occurring in the village. For ascending PS, the total displacement is highest in the areas where the ground slope is highest (Fig.~\ref{f_008}a). For descending PS we observe an opposite gradient: the total displacement is highest where the ground slope is lowest (Fig.~\ref{f_008}b).

\subsection{Geometric Configuration of Satellites Relative to the Target}\label{secPSgeometry}
\begin{figure}[!ht]
\centering
\includegraphics[width=0.7\textwidth]{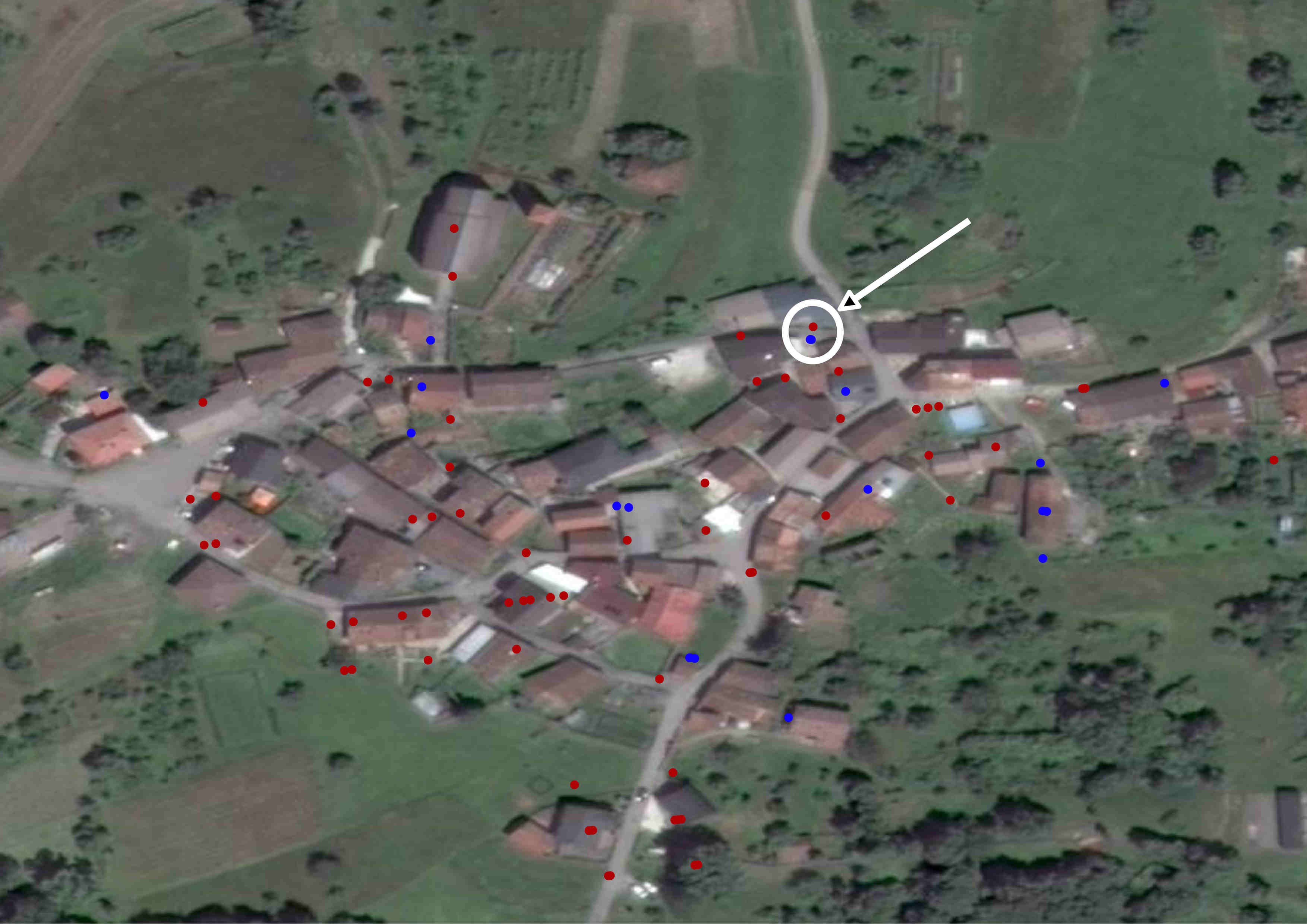}
\caption{Persistent scatterer both ascending and descending in the same map. The arrow highlights two PS, one in the ascending direction and one in the descending direction, located close to each other. The relative distance between the two PS is 3.8m.}
\label{f_009}
\end{figure}

Among the available data for each persistent scatterer, besides the geographical references of the PS, the details of the relative position of the satellites are available. Each satellite measures the projection of the PS displacement along the LOS. Since this is just a projection, it  provides only partial information about the real movement of the PS. To determine the displacement as a vector in space, two additional measurements are needed. To address this lack of information, we followed a conservative approach.
There is no PS that is observed both in ascending and in descending direction, that means that there is no ascending PS that coincides exactly with one of the descending PS in the village area. However, we observed that the time series of the PS are strongly correlated with each other. Therefore, the  points in the village, even if not identified as PS, are likely experiencing similar displacements to those measured for the PS. In other words, the village appears to be moving as a whole, in a consistent way. We can say that elastic movements between different PS, while they exist, are small and negligible compared to the collective movement identified by the PS, when the PS are not far from each other. We can assume that the displacement measurement of a PS in the ascending phase and the displacement measurement of a PS in the descending phase can be combined to estimate the displacement vector, even though the two PS are not perfectly coincident. In combining measurements generated from different orbit passes, we can neglect the time offset between samplings, if we consider the total displacement at the end of the sampling period. Fig.~\ref{f_009} shows an example of two PS in the ascending and descending directions, located close to each other, that could be used together to determine the displacement vector.

\begin{figure}[!ht]
\centering
\includegraphics[width=0.7\textwidth]{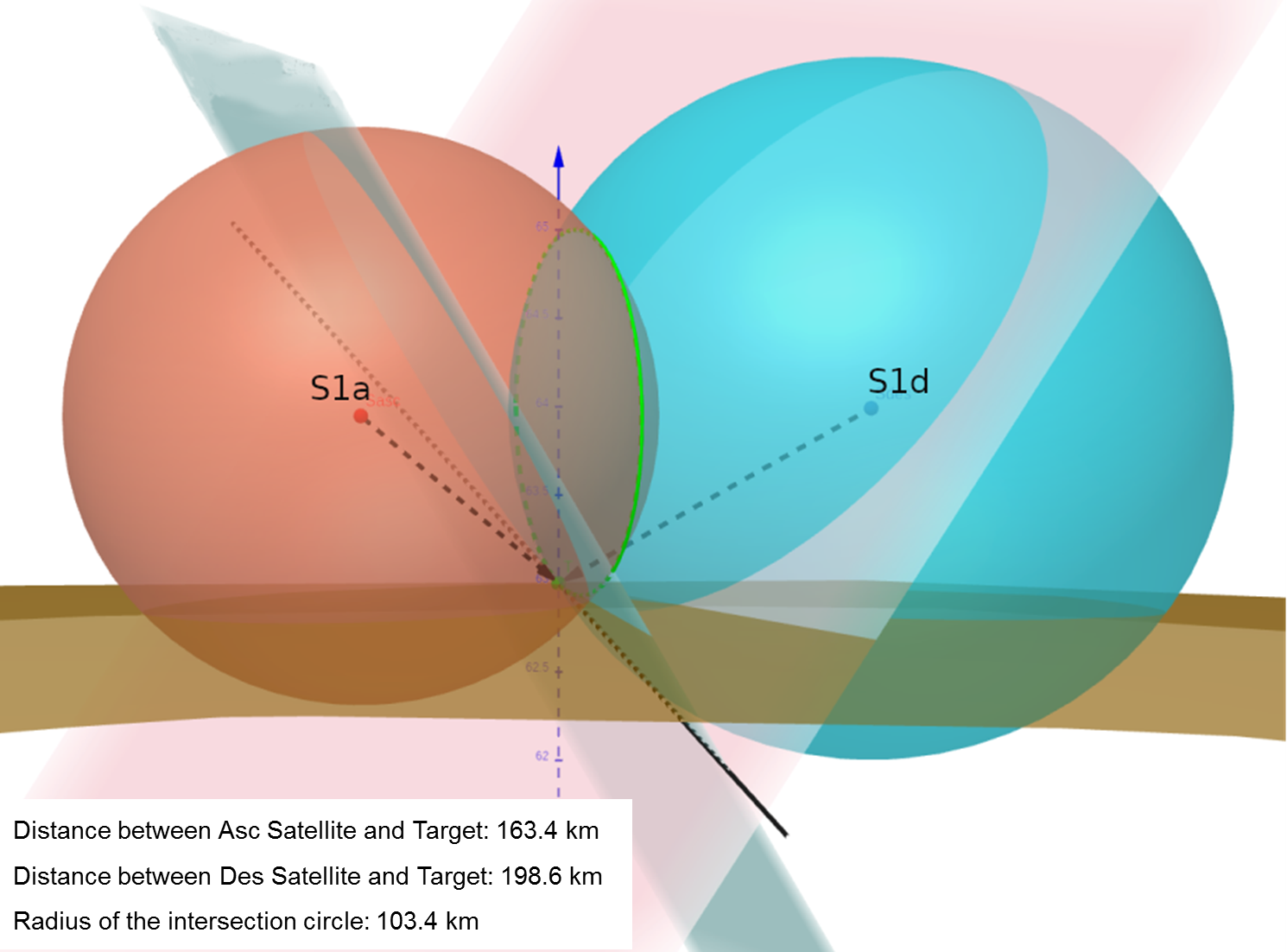}
\caption{Representation of the geometric configuration of the satellites relative to the target. The red sphere represents the locus of points at the same distance from the ascending satellite, while the blue sphere represents the locus of points at the same distance from the descending satellite. The green circle shows the intersection of the two spheres.}
\label{f_010}
\end{figure}

Suppose that the target is at a distance $r_a$ from the satellite in the ascending direction and $r_d$ from the satellite in the descending direction. In Fig.~\ref{f_010}, the red sphere represents the locus of the points located at a distance $r_a$ from the ascending satellite, while the blue sphere represents the locus of the points at a distance $r_d$ from the descending satellite. The intersection of these two spheres is shown as the green circle in the figure.
Since Sentinel-1 satellites travel at an average distance of about 100km from the earth surface and the displacements is of the order of the pixel area, we can approximate the spheres with planes without making significant errors.
\begin{figure}[ht!]
\centering
\includegraphics[width=0.5\textwidth]{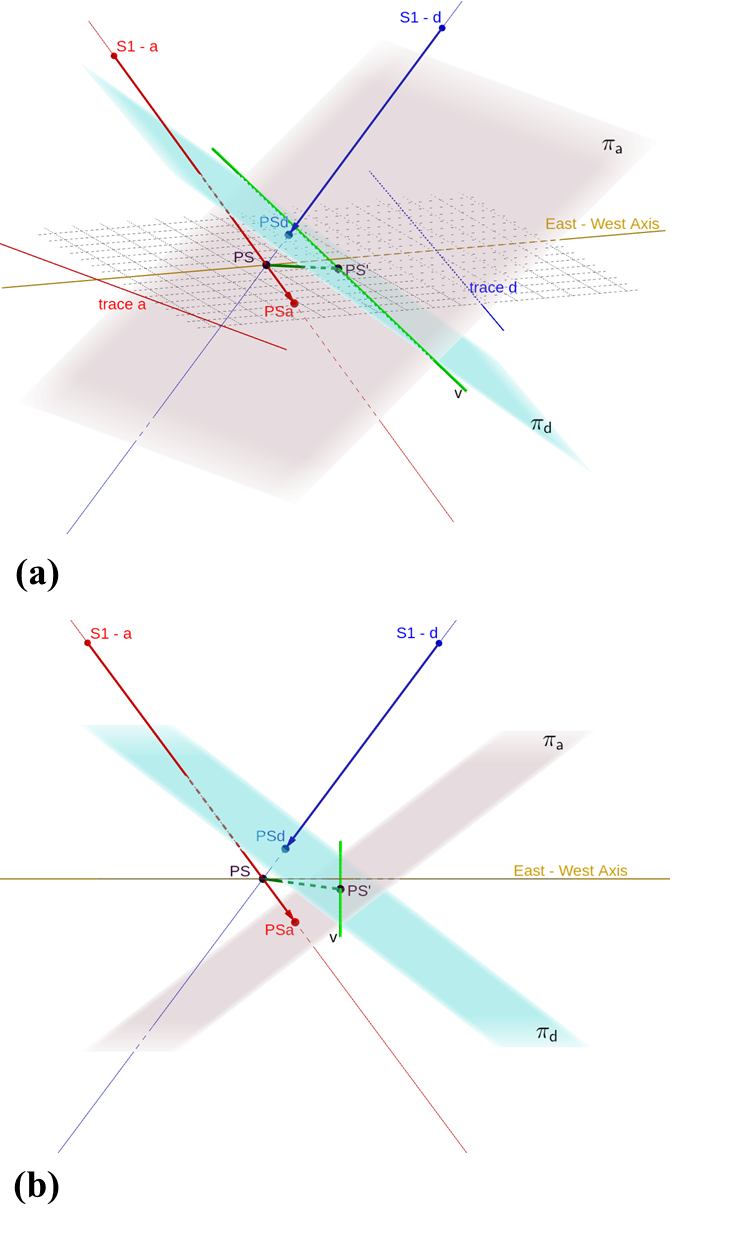}
\caption{Representation of the geometric configuration of the satellites relative to the target from two different perspectives. Near the PS, the spheres can be approximated by the red and blue planes, while the circle is approximated by the green line. The point $PS_a$ represents the way the ascending satellite measures the final position of the PS, while $PS_d$ represents the measurement from the descending satellite. The segment $PS-PS' $indicates the minimum possible displacement of the PS.}
\label{f_011}
\end{figure}
%\pagebreak 
Figure~\ref{f_011} shows the geometry of ascending and descending satellites ($S1_a$ and $S1_d$) relatively to the chosen PS. Satellite $S1_a$ measured PS distance, projected along the LOS, as if the PS moved away to $PS_a$. Observe that the PS could have moved on any of the point of a sphere (approximated with the plane $\pi_{a}$)
centered in $S1_a$ with radius equal to the distance between $S1_a$ and $PS_a$. Plane $\pi_{a}$ is considered as the locus of the points where PS has moved. On the other hand, satellite $S1_d$ measured that PS distance along LOS, decreased as if the PS moved to $PS_d$. By the same reasoning, we could state that $\pi_{d}$ is the locus of the points where PS has moved. The spheres are approximated as planes passing through the PS, and the green circle in Fig. \ref{f_010} is approximated as a line passing through the PS. At time 0, this line will pass through the initial position of the target (at the origin). At later times, as the target moves, the line will pass through the new position of the target.
Combining the two measurements, we can conclude that the PS has moved to one of the points of the straight-line $v$, intersection of $\pi_{a}$ and $\pi_{d}$. A third measurement (that would be able to remove the residual ambiguity, and uniquely determine one point in the line) is missing, but additional available info enables us to recover as much info as possible.
The green line represents the locus of the points to which the PS may have moved. 
The distance of the green line from the origin, represented by the segment $PS-PS'$ represents the \emph{minimum possible displacement} for the PS being analyzed. This is nothing but a lower bound for the absolute value of the  possible displacements of the PS.

\subsection{The Displacement field}
\begin{figure}[!ht]
\centering
\includegraphics[width=0.8\textwidth]{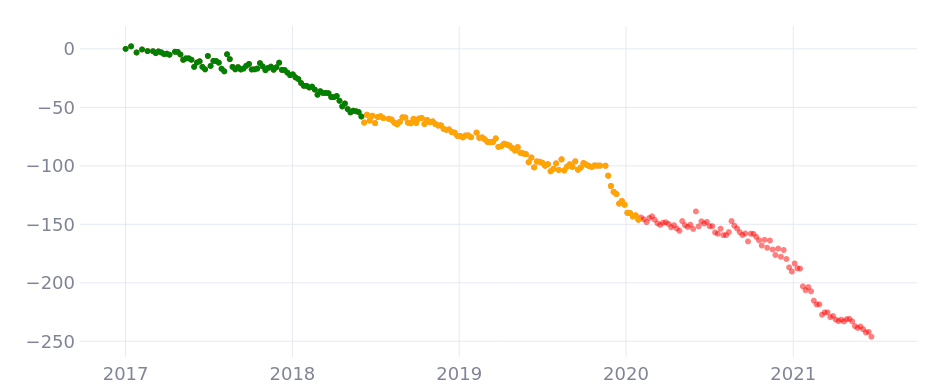}
\caption{Ascending PS LOS time series sections.}
\label{f_012}
\end{figure}

\begin{figure}[hb!]
\centering
\includegraphics[width=0.8\textwidth]{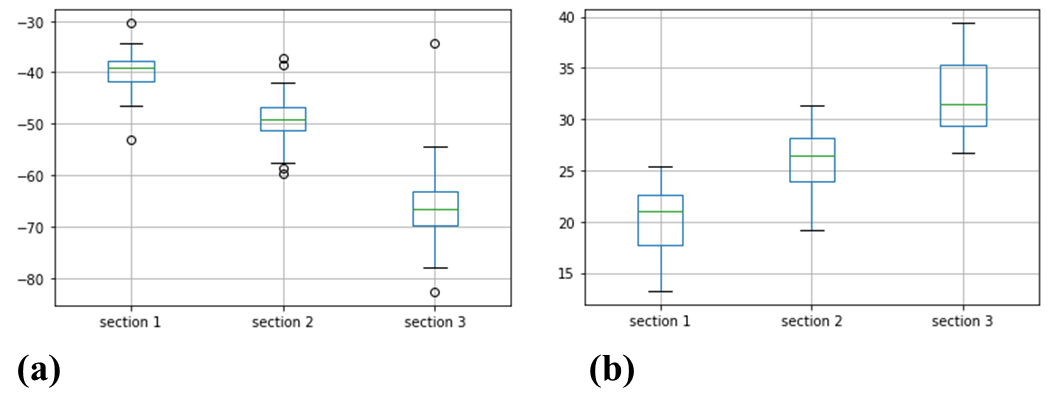}
\caption{PS LOS displacement rates (in mm/year): a) ascending PS; b) descending PS. The ascending PS LOS displacements exhibit a different trend compared to the descending PS LOS displacements.}
\label{f_013}
\end{figure}

By analyzing one of the ascending PS time series, it is possible to identify three contiguous main sections: the first one from the beginning of the observation to February 2018, the second one until February 2020 and a third one until the end of the series, as highlighted in Fig.~\ref{f_012}. In each section we can observe a regular trend followed by a steeper region.
Figure~\ref{f_013}a shows the statistics of the displacement rates in the three highlighted periods for all the ascending PS in the village. The same behaviour can be obviously observed in descending PS time series (Fig.~\ref{f_013}b).
However, by comparing ascending PS displacement rates with descending ones, we realize that they are increasing differently. This means that, according to the geometry of the system described in section \ref{secPSgeometry}, there is a change in the displacement vectors both in modulus and in direction: the two combined effects result in an increasing slope of the displacement vector.

Let us choose all the couples of ascending PS and descending PS whose relative distance is less than 20m: the two different measurements (ascending and descending LOS displacements) will be considered as arising from a single PS. Using the geometrical model described in section \ref{secPSgeometry}, the PS initial position is the origin of the coordinate system. The West-East axis is the $x$-axis, the North-South axis is the $y$-axis, the zenith-nadir axis the $z$-axis. If we define the coordinates of $S1_a$ and $S1_d$ as
\begin{align}
S1_{a} &= (a_{x}, a_{y}, a_{z}), \\
S1_{d} &= (d_{x}, d_{y}, d_{z}),
\end{align}
then we can write the equations of $\pi_{a}$ and $\pi_{d}$ as
\begin{align}
\pi_{a} &: a_{x}x + a_{y}y + a_{z}z - LOS_{a} = 0, \\
\pi_{d} &: d_{x}x + d_{y}y + d_{z}z - LOS_{d} = 0 ,
\end{align}
where $LOS_{a}$ an $LOS_{d}$ are the ascending and descending PS LOS displacements. The coordinates of the point $PS'$ are given by the intersection of $\pi_{a}$, $\pi_{d}$ and the plane passing through $S1_{a}$, $S1_{d}$ and $PS$ (the origin of the system). The vector defined by the points $PS$ and $PS'$ is the minimal-length vector among the possible vector displacements of PS. Lower bound on displacements yields conservative estimates, which, of course, suggests that the actual displacements are usually underestimated when considering only time series and displacements in a single direction. The mean of the values of the lower bound to the vector displacements shows an increment larger than 30\% compared to the single direction displacements (Tab.~\ref{tab6}).
\begin{table}[h!] 
\caption{Modulus of the PS total displacement vector: mean and standard deviation.\label{tab6}}
\newcolumntype{C}{>{\centering\arraybackslash}X}
\begin{tabularx}{\textwidth}{cCCC}
\toprule
\textbf{Vector}	& \textbf{Mean [mm]} & \textbf{Standard Dev [mm]} & \textbf{Time Interval} \\
\midrule
Modulus		    & -310  & 9  & 01/01/2017-21/06/2021\\
\bottomrule
\end{tabularx}
\end{table}

\begin{figure}[hb!]
\centering
\includegraphics[width=0.8\textwidth]{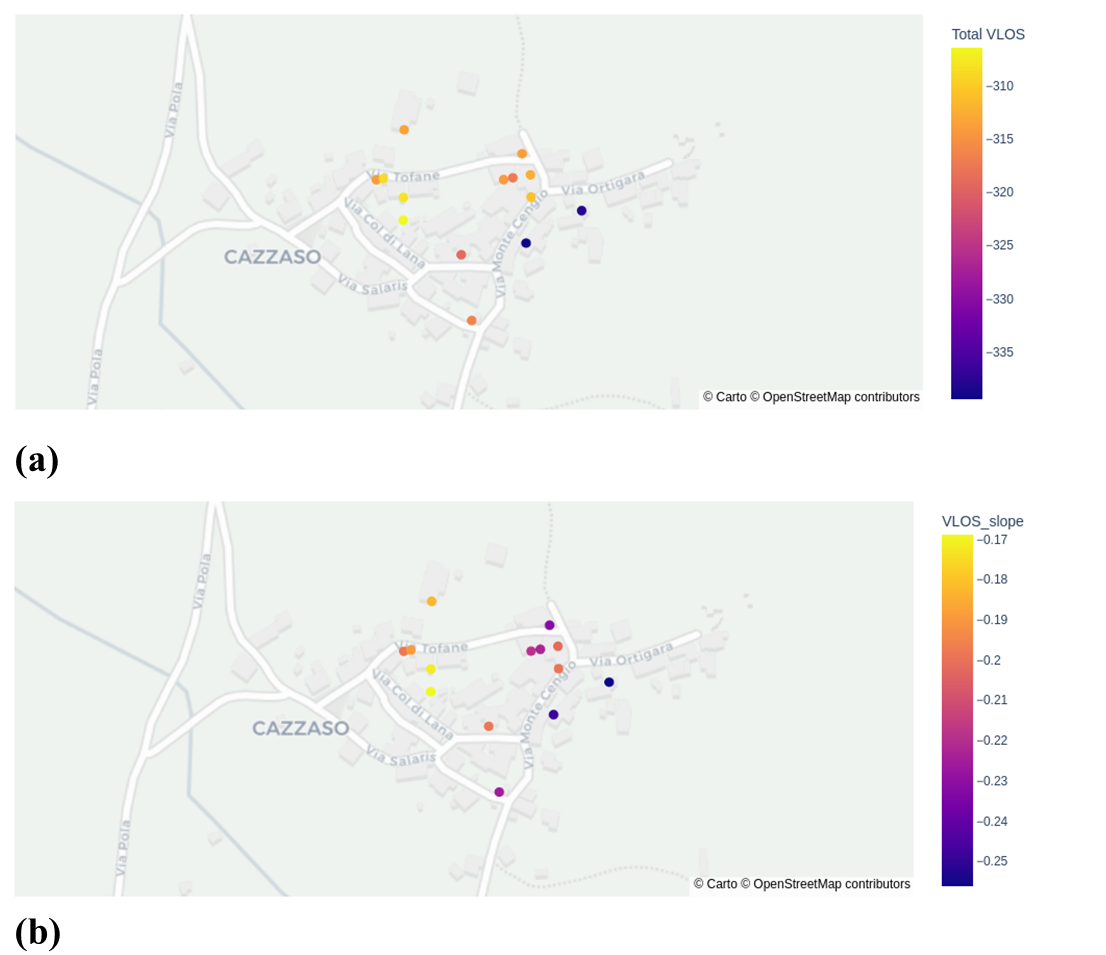}
\caption{Each point on the map of Cazzaso represents a pair of ascending and descending PS located close to each other, used to study the displacement vector. A gradient of the minimum possible displacement magnitude emerges across the area both in modulus (a) and in slope (b). The two gradients are coherent.}
\label{f_014}
\end{figure}

In Fig.~\ref{f_014} pairs of persistent scatterers that are close enough to each other are represented as a single point on the map of Cazzaso. The modulus and the vertical components of the lower bound to PS displacement vectors are displayed. On both maps, a coherent gradient can be observed: the modulus of the vector displacement is highest where the slope of the vector is highest. It is worth noticing that, while the measured displacement projections were less than 22 cm, the values of the minimum possible displacements reach up to about 34 cm.

\begin{figure}
\centering
\includegraphics[width=0.8\textwidth]{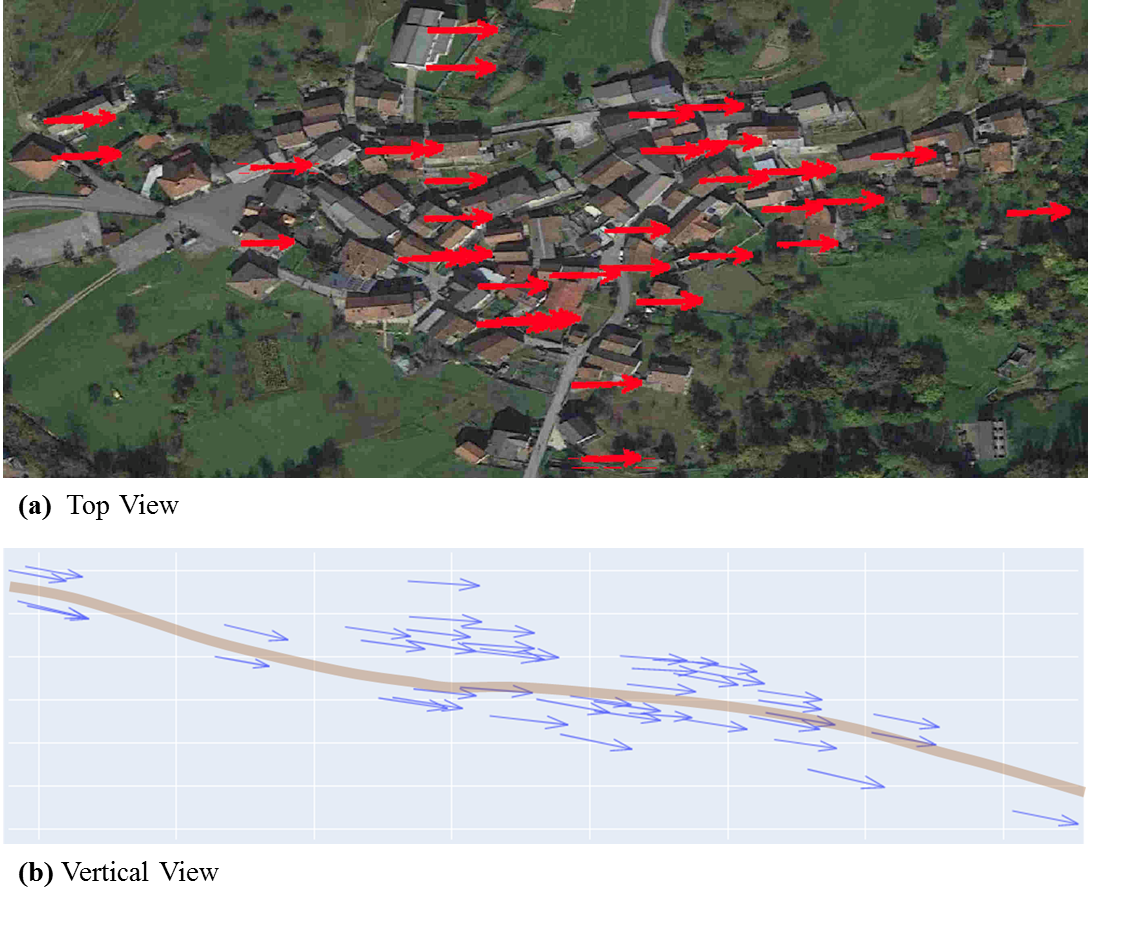}
\caption{Displacement field of the PS in the village of Cazzaso.}
\label{f_550}
\end{figure}

To better understand the direction of the displacements of the persistent scatterers, the terrain morphology was analyzed, including its 3D structure, to understand the layout of the village relative to the slopes of the land (Fig.~\ref{f_003}). By using the altitude measurements of the PS, it was possible to calculate the average slope of the terrain in the whole area of the village. This analysis enabled us to reveal how the village is positioned on the mountainside and how the terrain's slope influences its stability. 
This result is coherent with the test site ground profile. 
Combining the terrain study with the analysis of the PS displacements following the previously explained procedure, it was possible to determine the velocity field of the PS across the village: see Fig.~\ref{f_550}, which is central to this work. The displacement field is represented in two different projections: in Fig.~\ref{f_550}a it is shown from above on the horizontal plane, and in Fig.~\ref{f_550}b it is shown on the vertical plane.
The results clearly show that the village of Cazzaso is sliding down \emph{along} the mountain slope. Technically, the velocity field is \emph{tangent} to the terrain. Observe that this conclusion is obtained from a careful analysis of the data and is not put ``by hand''.
In conclusion, the entire area is affected by a continuous movement, caused by the combination of the steep slope and other external factors like gravity and weather conditions. 

%%%%%%%%%%%%%%%%%%%%%%%%%%%%%%%%%%%%%%%%%%
\section{Discussion and conclusions}
The procedure applied to the village of Cazzaso successfully highlighted geological dynamics. The proposed procedure was applied as a post-processing step to the data, regardless of the specific processing chain used to generate them. It does not require any additional processing but operates directly on the time series, which have already been corrected and stripped of various error components, without requiring any intervention in the previous processing steps.  

Phase unwrapping enhancement played a key role. Only after improving the phase unwrapping using the proposed methods was it possible to identify deformation gradients on the Earth's surface. By analyzing the satellite geometry in ascending and descending phases relative to the target, it was possible to estimate a lower bound for the displacement magnitude and determine the geometrical locus of all possible displacement locations. Importantly, this study was conducted solely using satellite data, without requiring additional measurements on site. In the absence of a measured third coordinate, data from ascending and descending satellite  were integrated with a general analysis of the terrain's geometry and morphology. This integration provided insights into the displacement geometry and allowed for the determination of the velocity field.  

From the time series of individual persistent scatterers unwrapped before enhancement, total displacements along a single component (ascending or descending) were initially found to be less than 20 cm. However, after applying the proposed corrections and combining ascending and descending data for the same PS, a lower bound of 34 cm was identified for the total displacement. This represents an increase of approximately 70\%. We emphasize that this estimate is conservative, as discussed in the text. This significant change highlights the necessity of these improvements. The proposed methodology automates this critical enhancement process, significantly reducing the amount of manual intervention, though not eliminating it entirely. This allows analysts to focus more on the global aspects of the phenomenon being studied, ensuring greater accuracy and reliability in the results, while maintaining efficiency and minimizing the potential for human error. Further research could extend these findings to other case studies, refining the methodologies and exploring their broader applicability.

%%%%%%%%%%%%%%%%%%%%%%%%%%%%%%%%%%%%%%%%%%
\vspace{6pt}

%%%%%%%%%%%%%%%%%%%%%%%%%%%%%%%%%%%%%%%%%%
\authorcontributions{GB and RN designed the study; GB and SP performed analyses and wrote the paper; SP supervised the development of the model; CL contributed to the supervision of the research project. GB, RN, PF, LG, FVP, CL, and SP interpreted the results, revised the text and approved the final version of the paper.}

\funding{We acknowledge support from the PNRR MUR project CN00000013-``Italian National Centre on HPC, Big Data and Quantum Computing’’. GB acknowledges the support from the European Space Agency (ESA), Co-Sponsored Research contract on "Quantum computing for ground motion", Contract No. 4000139394/22/I-DT-lr. GB, PF, FVP and SP acknowledge the support from Istituto Nazionale di Fisica Nucleare (INFN) through the project ‘QUANTUM’. PF, SP and FVP acknowledge the support from PNRR MUR project PE0000023-NQSTI.  We acknowledge financial support under the PNRR M4.C2.1.1. MUR PRIN 2022, funded by the European Union – NextGenerationEU – Project Title ``QUEXO'' – CUP:D53D23002850006.
We acknowledge support from the Italian funding within the ``Budget MUR - Dipartimenti di Eccellenza 2023–2027'' - Quantum Sensing and Modelling for One-Health (QuaSiModO).
}

\acknowledgments{Code testing were performed on the IT resources hosted at ReCaS data center (project financed by the italian MIUR (PONa3\_00052, Avviso 254/Ric.)).
The view expressed herein can in no way be taken to reflect the official opinion of the European Space Agency.}

\conflictsofinterest{The authors declare no conflict of interest.}

\abbreviations{Abbreviations}{
The following abbreviations are used in this manuscript:\\

\noindent 
\begin{tabular}{@{}ll}
DInSAR & Differential SAR Interferometry\\
EEA & European Environment Agency \\
LOS & Line of sight\\
MTInSAR & Multi-Temporal SAR Interferometry\\
PS & persistent Scatterer\\
VLOS & Velocity along the satellite's line of sight\\

\end{tabular}
}

%%%%%%%%%%%%%%%%%%%%%%%%%%%%%%%%%%%%%%%%%%
\begin{adjustwidth}{-\extralength}{0cm}

\reftitle{References}

\end{adjustwidth}
\end{document}